# Phosphorus Nanotubes from Chemical Cleavage

Romakanta Bhattarai[1] and Xiao Shen[2*]

[1]Department of Physics, Applied Physics, and Astronomy, Rensselaer Polytechnic Institute

Troy, NY, 12180

[2]Department of Physics and Materials Science, The University of Memphis

Memphis, TN, 38152

Corresponding author: xshen1@memphis.edu

**Abstract**

We propose a strategy to make phosphorus nanotubes from two well-known phosphorus allotropes: violet phosphorus and fibrous red phosphorus. First-principles calculations show that doping with sulfur dissociates the covalent bonds between tubular phosphorus structures that form bilayers in these allotropes, resulting in free-standing 1D nanotubes. Due to the substitutional nature of the sulfur dopant, the resulting 1D structure is linear, unlike the helical ring structure studied previously. The sulfur sites are situated periodically along the 1D nanotubes and can be further functionalized. Our results show that the S-doped phosphorus nanotube can sustain a tensile strain of up to 18%. The strain also substantially modifies the electronic band gap and the effective mass of carriers. Calculations using the many-body Green's functions (GW) and the Bethe-Salpeter equation (BSE) approaches reveal a large exciton binding energy of 1.57 eV. The one-dimensional nature, linearity, functionalizability, mechanical flexibility, tunability of electronic properties, and large exciton binding energy make this material interesting for applications in optoelectronic devices, solar cells, chemical sensors, and quantum computing.

**Introduction**

After the successful synthesis of single atomic layer graphene in 2004,[1] two-dimensional (2D) materials have been at the center of attention, leading to the discovery of many other atomically thin van der Waals (vdW) materials such as transition metal dichalcogenides (TMDs)[2] and hexagonal boron nitrides (hBNs).[3] Graphene has large carrier mobility, but its gapless nature makes applications in electronic devices challenging.[4] On the other hand, TMDs have non-zero band gaps, but their carrier mobilities are not high, which also hinders them from a wide range of applications.[5]

Phosphorus allotropes have gained significant attention as vdW materials. In 2014, Liu et al. successfully synthesized black phosphorene (α-P) from bulk black phosphorus. The newly discovered α-P has notable properties, such as high stability, a non-zero direct band gap, and high carrier mobility, making it a promising material for applications in optoelectronic devices. In 2014, the Tomanek group investigated a new phase of 2D blue phosphorus (β-P), which has an indirect band gap of ~2 eV. In 2016, Schusteritsch et al. investigated 2D violet phosphorus, also called Hittorf's phosphorus, which has a wide band gap (~2.5 eV) and high anisotropic carrier mobility. In 2017, Han et al. theoretically predicted green phosphorus (λ-P) to be more energetically stable than β-P. Several other phosphorus allotropes in 2D forms have been investigated through theoretical calculations and experimental observations. Examples include γ-P, θ-P, ε-P, ψ-P, δ-P, η-P, ζ-P, and φ-P. These allotropes possess a variety of electronic and optical properties.

In 2016, the Tomanek group investigated a stable phase of 1D phosphorus allotrope using genetic algorithm and density functional theory calculations. This 1D phosphorus features a repeating $P_2$-$P_8$ unit cell, forming a stable helical coil with a radius of 2.4 nm.[15] In this work, we propose that a new linear 1D phosphorus structure can be made by chemical cleavage from two well-known phosphorus allotropes: violet phosphorus and fibrous red phosphorus. Calculations show that the substitutional doping of a pair of sulfur atoms dissociates the covalent bonding between the two phosphorus layers in these allotropes, resulting in free-standing nanotubes. Compared to the helical coil structure,[15] the linearity of the nanotube is due to the additional sulfur atoms that straighten the chain structure. We investigate the resulting S-doped phosphorus nanotube's mechanical, electronic, and transport properties when a tensile strain is applied. We also calculate the frequency-dependent dielectric function using many-body Green's functions (GW) and the Bethe-Salpeter equation (BSE) approaches, revealing an exciton binding energy among the largest. The Raman spectrum is also calculated to facilitate experimental explorations.

**Methods**

First-principles calculations are performed using the VASP package,[16] where the interaction between the electrons and ions is described by the pseudopotentials constructed under the projector augmented wave (PAW) method.[17] The GGA-PBE exchange-correlation functional is used.[18] A plane wave basis set of energy 500 eV is used to expand the electronic wave function. The break condition for electronic relaxation is set to $10^{-8}$ eV between consecutive steps. The structure is relaxed until the force on each ion is less than 0.001 eV/Å. Gamma-centered k-point grids of 5 × 5 × 1 and 15 × 1 × 1 are used for the integration of Brillouin zones for the 2D and 1D structures, respectively. To avoid the vdW interactions between any periodic replicas, sufficiently large vacuum spaces of 15 and 28 Å are used along the y- and z-axes, respectively. The frequency-dependent dielectric function is calculated by the many-body Green's functions (GW)[19–21] and Bethe Salpeter equation (BSE)[22–24] methods. The Raman spectrum and the corresponding vibrational modes are calculated by the density functional perturbation theory[25] using the Quantum Espresso package,[26] which uses the norm-conserving pseudopotentials generated via the Rappe-Rabe-Kaxiras-Joannopoulos (RRKJ)[27] scheme and the plane wave basis set with the energy cutoff of 80 Rydberg (Ry) are used. The total energy and force convergence criteria during ionic relaxation are set to $10^{-6}$ Ry and $10^{-4}$ Ry/Bohr, respectively.

**Results and Discussion**

Figure 1 shows the crystal structures of two allotropes of phosphorus: violet (a) and fibrous red (b) phosphorus. Both allotropes feature bi-layer structures, with each single layer composed of long tubular structures of phosphorus (P) atoms with repeating $P_2$-$P_9$-$P_2$-$P_8$ units. These two layers are connected to each other by covalent bonds formed between pairs of P atoms, each from a $P_9$ cluster on the surface of the tubular structures, with bond distances of 2.210 Å in violet phosphorus and 2.246 Å in fibrous red phosphorus. In violet phosphorus, the tubular structures in the two single layers are perpendicular to each other. In fibrous red phosphorus, however, the tubular structures in the two single layers are aligned in parallel. The P atoms connecting the single layers are three-folded, with two intra-layer P-P bonds and one inter-layer P-P bond. We expect that modifying these connecting P atoms can alter the inter-layer bonds and potentially eliminate them.

Sulfur atoms are a natural choice for eliminating the inter-layer bonds, as they usually take a 2-fold bonding structure. Their similar size to P atoms also facilitates substitutional doping. The question is whether sulfur atoms prefer to substitute the linking P atom or other P atoms that do not participate in the inter-layer bonds. To test this, we first substitute one P atom into the 2D violet phosphorus and compare the energies of three possible sites: the linking P site, a non-linking P site on the inner side of the double layer, and a position on the outer side. The results show that doping one S atom at the linking P site is more energetically favorable than the latter two cases by 0.23 and 0.24 eV, respectively. Next, we substitute two P with two S atoms. We consider 23 pairs of possible P sites, including those in the same layer and one site in each layer. Our calculations show that substituting the P-P pair connecting the two layers has lower energy than any other combinations by at least 0.21 eV. In this most energetically favorable configuration, each S atom forms a 2-fold bonding structure with two intra-layer S-P bonds, as expected. Thus, the original inter-layer covalent bond is eliminated. This results in two disconnected S-doped phosphorus layers separated by a distance of 3.595 Å, as shown in Figure 1(c), with each layer consisting of aligned S-doped phosphorus nanotubes. For fibrous red phosphorus, the most energetically favorable S-doping sites are the same as those for violet phosphorus. The vertical distance between the two separated layers is 3.152 Å, as shown in Figure 1(d). The large separations indicate that the resulting two single layers now have only a weak van der Waals interaction.

From both violet and fibrous red phosphorus, we obtain the same nanotube structure. Figure 1(e-g) shows different views of the lattice structure of the resulting nanotube. The unit cell consists of 21 atoms with an S to P ratio of 1:20. The lattice vector of the optimized structure is 13.009 Å along the direction of the nanotube, which is the horizontal direction in Figures 1(e) and 1(g). The top view in Figure 1(e) clearly shows that the structure has a repetitive unit cell of $P_2$-$P_8S_1$-$P_2$-$P_8$. Compared to the helical coil structure discovered by the Tomanek group in 2016, which has 10 atoms in a unit cell with a repetitive unit of $P_2$-$P_8$, the addition of an S atom in one of the two $P_8$ units doubles the unit cell size in the present structure. The additional S atom straightens the structure, akin to the original tubular structure in violet and fibrous red phosphorus. The tubular nature is clearly demonstrated in the front view of the structure, shown in Figure 1(f). The side view, shown in Figure 1(g), demonstrates three types of ring structures: one hexagon on the left, one incomplete hexagon in the middle where one P atom is missing, and one pentagon on the right. (Note that not all the P atoms are displayed in this cross-sectional view; that is, 8 P atoms are hidden.) Importantly, the sulfur sites are situated periodically in a configuration that can be easily accessed, facilitating further functionalization and manipulation of the as-cleaved nanotubes.

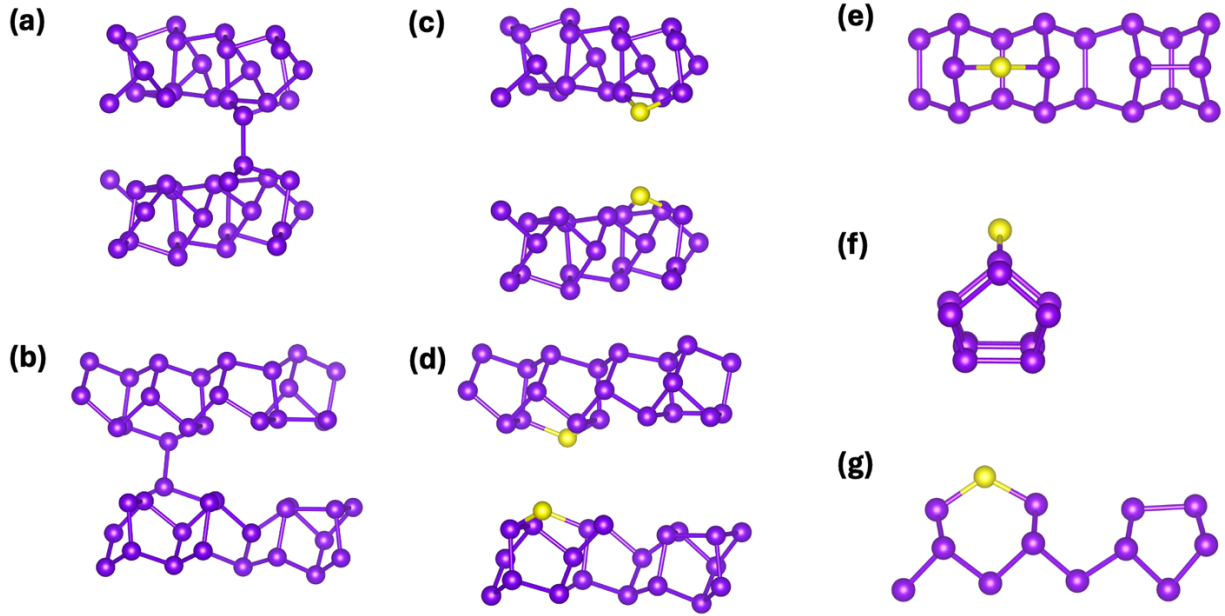

Figure 1. Crystal structures of (a) violet phosphorus, (b) fibrous red phosphorus, (c) sulfur-doped violet phosphorus, and (d) sulfur-doped fibrous red phosphorus. Panels (e) – (g) represent the S-doped phosphorous nanotube in the top, front, and side views. Phosphorus and sulfur atoms are shown in violet and yellow, respectively.

Next, we investigate the properties of the S-doped phosphorus nanotube in terms of its response to mechanical strain. A set of tensile strains is applied along the nanotube, and the corresponding forces are computed and shown in Figure 2(a). We find that the 1D nanotube can sustain a mechanical strain of up to 18%, which is the critical strain for this material. If the strain is further increased, the structure undergoes a phase transition and eventually collapses, as shown by the sudden drop in tensile force beyond the 18% limit. The force corresponding to maximum strain is 5.155 nN for a single nanotube.

The electronic properties of nanotubes exhibit a large response to the applied strain. Figure 2(b) shows the change in the band gap. At 0% strain or the unstrained case, the band gap is 2.18 eV. As the strain increases, the gap starts decreasing, eventually reducing to 0.253 eV at the critical strain value of 18%. The large change in the band gap is accompanied by a significant change in the effective masses of carriers, as shown in Figure 2(c). In the unstrained case, effective masses for electrons and holes are 0.903 $m_0$ and 0.748 $m_0$, respectively. When strain is applied, $m_e$ first decreases and reaches a minimal value of 0.152 $m_0$ at 4% strain, then increases to 0.362 $m_0$ at 18%. Meanwhile, $m_h$ always decreases with strain and reaches 0.177 $m_0$ at 18%. These low values are comparable to or smaller than the effective masses in single-walled carbon nanotubes.[28] Overall, these results suggest that it is possible to use mechanical strain to achieve low effective masses and, hence, high mobility in this nanotube.

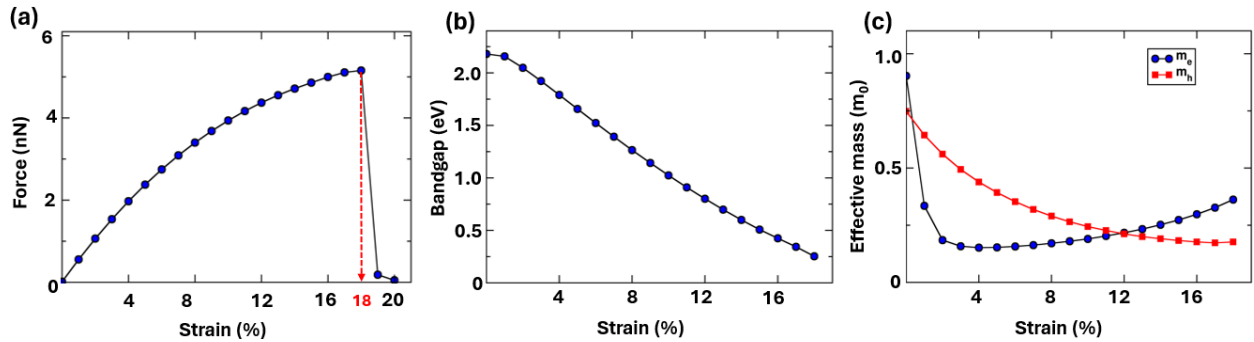

Figure 2. (a) Force-strain curve of a S-doped phosphorous nanotube showing the critical strain. (b) Effect of tensile strain on the electronic band gap. (c) Effect of tensile strain on the effective masses of electrons and holes.

The large changes in the band gaps and effective masses due to tensile strain are associated with the changes in band structures shown in Figure 3. In the unstrained case, the nanotube has an indirect band gap of 2.18 eV, with the maximum of the valence band (VBM) lying at the X point in the Brillouin zone and the minimum of the conduction band (CBM) lying between the Γ and X, represented by two black circles in Figure 3(a). The energy bands closest to the Fermi level on either side (below and above) are mostly flat with very low curvatures. Thus, the larger effective masses in Figure 2(c) result from the flat bands near the Fermi level. When the strain is applied, the position of the VBM remains unchanged, whereas the CBM shifts to the Γ point at 5% (Figure 3(b)) and remains unchanged thereafter. Meanwhile, the curvatures of those energy bands change with the strain, thus altering the carriers' effective masses.

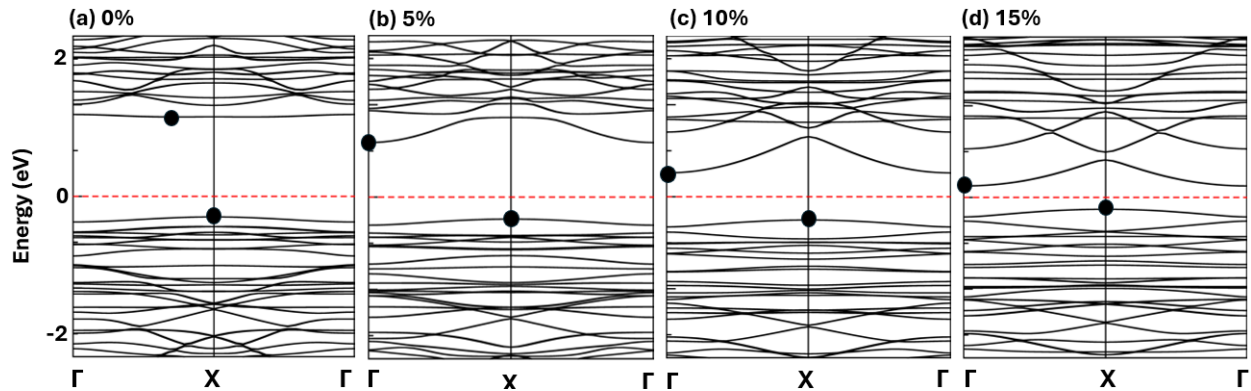

Figure 3. Band structure of S-dope phosphorus nanotube under different strains.

To further understand the effect of applied strain on the electronic structure, we plot the partial charge densities of the relevant bands at different strain values in Figure 4. The upper row (a-d) represents the partial charge densities at the lowest conduction band, and the lower row (e-h) represents those of the highest valence band. In an unstrained case, the lowest conduction band, shown in Figure 4(a), is mainly composed of the orbitals of the sulfur dopant atom and the phosphorus atoms it is bonded with. Small contributions also come from the phosphorus atoms on the opposite sides of the hexagonal rings (marked as '1'). The overall charge density of the lowest conduction band is largely localized in the $P_8S_1$ unit, shown as the hexagonal ring in the side view, mainly on the S side. The localized nature of the state results in a flat band dispersion. When strain is applied, the contribution from sulfur starts decreasing and eventually vanishes as the strain increases. However, the contribution from other phosphorus atoms on the opposite side of the nanotube, notably those in the $P_2$ units, increases with the strain. Gradually, an extended state is formed, as shown in Figure 4(b-d), which leads to a larger dispersion of the lowest conduction band in Figure 3(b-d).

Contrary to the case of the lowest conduction band, in an unstrained case, we don't see any contribution to the highest valence band from the sulfur atom, which is marked as '2' in Figure 4(e). The main contributions to the highest valence band come from phosphorus atoms in the $P_8$ unit (the pentagon in the side view), which are marked as '3' in Figure 4(e). Therefore, the highest valence band is also localized; thus, it has a small dispersion. When strain is applied, the P atoms in the $P_2$ units, which are bonded to the aforementioned phosphorus atoms, start to contribute, as shown in Figure 4(f-h). Meanwhile, the sulfur atom begins to contribute slightly, as indicated by the small charge densities in Figure 4(f-h). This, again, leads to the formation of an extended state, resulting in a larger dispersion. Overall, the $P_8S_1$ unit (hexagonal ring) is the main contributor to the lowest conduction band; the $P_8$ unit (the pentagonal ring) is mainly responsible for the highest valence band; and the $P_2$ unit contributes to both bands under tensile strain. The sensitivity of the electronic structure to the strain reflects the change in the nature of the states in the conduction and valence bands, especially regarding their localization and extension.

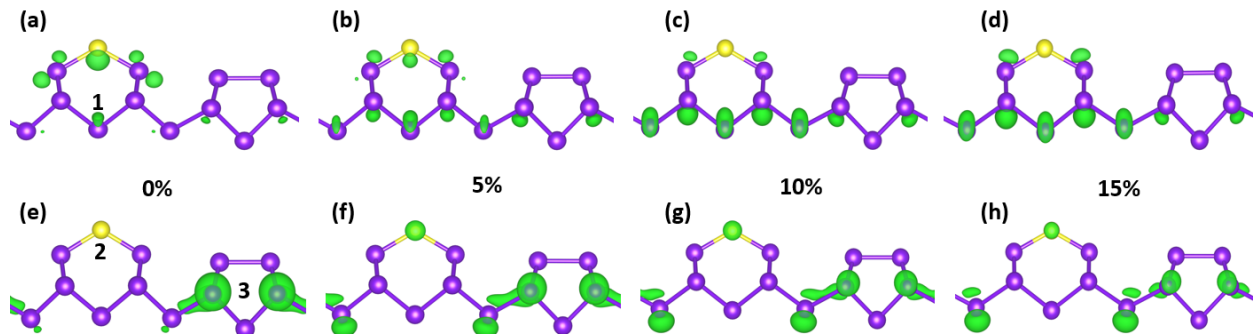

Figure 4. Partial charge densities at 0%, 5%, 10%, and 15% tensile strain of S-doped phosphorus nanotube. (a-d) CBM and (e-h) VBM. The isosurfaces of the charge density are shown in green.

We explore the optical properties of the S-doped phosphorus nanotube using the many-body Green's function (GW) method, which does not include the excitonic effect, and the Bethe-Salpeter equation (BSE) method, which includes the excitonic effect. A plot of the imaginary part of the frequency-dependent dielectric function along the x-axis, i.e., Im($\varepsilon_{xx}$), as a function of incident photon energy is shown in Figure 5(a). From the GW calculations, we find the quasiparticle band gap of 4.10 eV, and from the BSE calculations, we find the excitonic band gap of 2.53 eV. The exciton binding energy thus obtained is 1.57 eV, which is among the highest. The large binding energy results from the combination of the low effective masses of carriers and the limited dielectric screening in a one-dimensional system. The magnitude suggests that the exciton might be Frenkel in nature. The large exciton binding energy implies that the electrons and holes are very strongly bound to each other in this material. Since the energy corresponding to the thermal fluctuations (26 meV) is unlikely to dissociate those excitons, they should be very stable even at room temperature, resulting in a longer lifetime. The long lifetime of excitons is particularly desirable for quantum computing applications.[29] Easy access to the S sites for functionalization might facilitate tuning the excitonic properties. Furthermore, the sensitivity of the electronic structure to tensile strain might enable mechanical control of the spatial confinement of the excitons. This material might also be interesting for other applications in optoelectronic devices such as light-emitting diodes, solar cells, and photodetectors.

To assist experimental exploration, we calculate the Raman spectrum of the S-doped phosphorus nanotube using density functional perturbation theory. The result is shown in Figure 5(b), along with the corresponding vibrational modes. Four major Raman-active peaks are observed at 338, 372, 379, and 440 cm$^{-1}$. The vibrational modes show that the Raman peaks at 338 and 379 cm$^{-1}$ correspond to the out-of-plane vibration of P atoms, while the modes at 372 and 440 cm$^{-1}$ correspond to the in-plane vibration of P atoms. In addition, two minor Raman peaks at 203 and 324 cm$^{-1}$ are also observed.

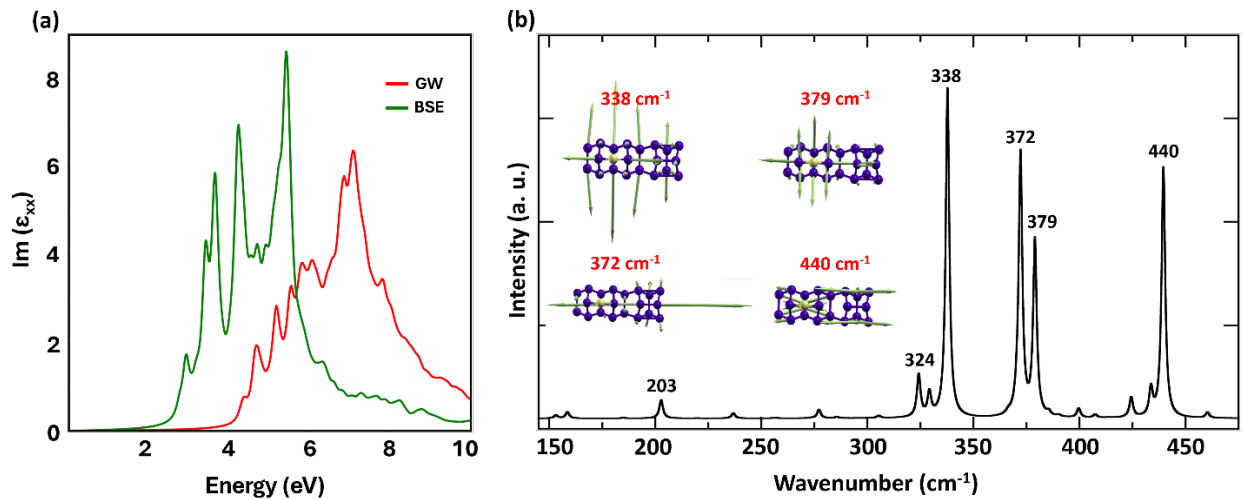

Figure 5. (a) Optical spectra of a S-doped phosphorus nanotube using the GW and BSE methods. (b) Calculated Raman spectrum of the nanotube, showing the four major Raman peaks and their respective phonon vibration modes, along with two minor Raman peaks.

**Conclusions**

In summary, we propose a phosphorus nanotube structure created by chemical cleavage from two well-known phosphorus allotropes: violet phosphorus and fibrous red phosphorus. Calculations show that the most energetically favorable sites for sulfur incorporation are the phosphorus atoms that form covalent bonds connecting two phosphorus layers. The doping dissociates the inter-layer bonds, resulting in free-standing S-doped phosphorus nanotubes. The linearity of this structure is ensured by the presence of additional sulfur atoms that form hexagonal rings with five other P atoms in the structure, unlike the previously studied helical coil phase of 1D phosphorus. We find that 1D phosphorus can sustain a tensile strain of up to 18%. The calculated dielectric functions indicate an extremely large exciton binding energy of 1.57 eV, among the highest in any material. Raman spectrum calculations are also performed to facilitate experimental observations. The large exciton binding energy and tunable electronic properties enable this material to have potential applications in optoelectronic devices, solar cells, chemical sensors, and quantum computing.

**Author contributions**

R.B. and X. S. conceived the idea of the project. R. B. performed the calculations, analyzed the results, and wrote the manuscript. X. S. assisted in the data analysis and co-wrote the manuscript.

**Competing interests**

The authors declare no financial competing interests.

**Acknowledgments**

The computational resources are provided by the High-Performance Computing Center (HPCC) at the University of Memphis.